\documentclass{iau} 
\usepackage{graphicx}
\usepackage{amsmath}

\title[JD 11.~~Twisted Magnetospheres] 
{Force-free and twisted, relativistic neutron star magnetospheres}

\author[Dimitris Ntotsikas \& K.N.Gourgouliatos]   
{Dimitris Ntotsikas$^1$
 \and Konstantinos N. Gourgouliatos$^1$}

\affiliation{$^1$University Of Patras, Department Of Physics, Patras, 26504, Greece \\
www.astro.upatras.gr \\ email: {\tt d.ntotsikas@upnet.gr}}
\pubyear{2022}
\volume{363}  
\jname{Neutron Star Astrophysics at the Crossroads: \\Magnetars and the Multimessenger Revolution.}
\editors{Prof. E. Troja \& Prof M. Baring, eds.}
\begin{document}

\maketitle

\begin{abstract}
In this poster we present the structure of an axisymmetric, force-free magnetosphere of a twisted, aligned rotating dipole within a corotating plasma-filled magnetosphere. We explore various profiles for the twist. We find that as the current increases more field lines cross the light cylinder leading to more efficient spin-down. Moreover, we notice that the twist cannot be increased indefinitely and after a finite twist of about $\pi/2$ the field becomes approximately radial. This could have implications for torque variations of magnetars related to outbursts.
\end{abstract}

Magnetar fields are believed to be strongly twisted due to shearing of the crust by internal magnetic stresses. A twisted magnetic field tends to inflate in the radial direction (\cite{1994MNRAS.267..146L, 2013ApJ...774...92P}). We find such solutions by applying the method of simultaneous relaxation (\cite{contopoulos1999axisymmetric}) for the magnetic field inside and outside the light cylinder, demanding a smooth crossing of the field lines on the light cylinder, using a suitably modified version of a previously developed numerical code (\cite{gourgouliatos2019coupled}). The pulsar equation is the following (\cite{goldreich1969pulsar}): 
\begin{equation*}
\tag{1}
     (1 - R^2)\Big(\frac{\partial^2\Psi}{\partial R^2} - \frac{1}{R}\frac{\partial\Psi}{\partial R} +  \frac{\partial^2\Psi}{\partial z^2}\Big) -2R\frac{\partial\Psi}{\partial R} = I(\Psi)\frac{dI(\Psi)}{d\Psi}
     \end{equation*}
where $\Psi$ is the magnetic flux function, $I$ is the electric current distribution and the light cylinder is at $R=1$. We assume that the geometry of the generated magnetic field is that of an aligned rotating star. For closed field lines the electric current distribution, $I$, is given by the expression:\quad $I(\Psi) = \kappa(\Psi(R_{in},Z_{in}) - \Psi_0 )^n  $ 
where $\kappa , n $ are free parameters and $R_{in}, Z_{in}$ describe the domain of the magnetosphere within the cylinder and $ \Psi_0$ is the magnetic flux function of the last closed magnetic field line. The present model cannot be modified for the case of an oblique rotator as the computational code does not take into account the time evolution of the system. 
There is great freedom on the relation between $I$ and $\Psi$, the one we follow comes from the relationship (\cite{vigano2013magnetic}, \cite{10.1093/mnras/stz2438}):
\begin{equation}
\tag{2}
I(\Psi) =\int_0^{\Psi}\alpha(\Psi')d\Psi' , \quad \alpha = \frac{\kappa}{R_{\star}}\left(\frac{\Psi}{\Psi_0}\right)^q
\end{equation}
with $\kappa$ being the dimensionless parameter  related to the twist, and $q$ a free parameter in correspondence with the parameter $n$, $n=q+1$.

\begin{figure*}
    \centering
    (a) \includegraphics[width=0.40\textwidth]{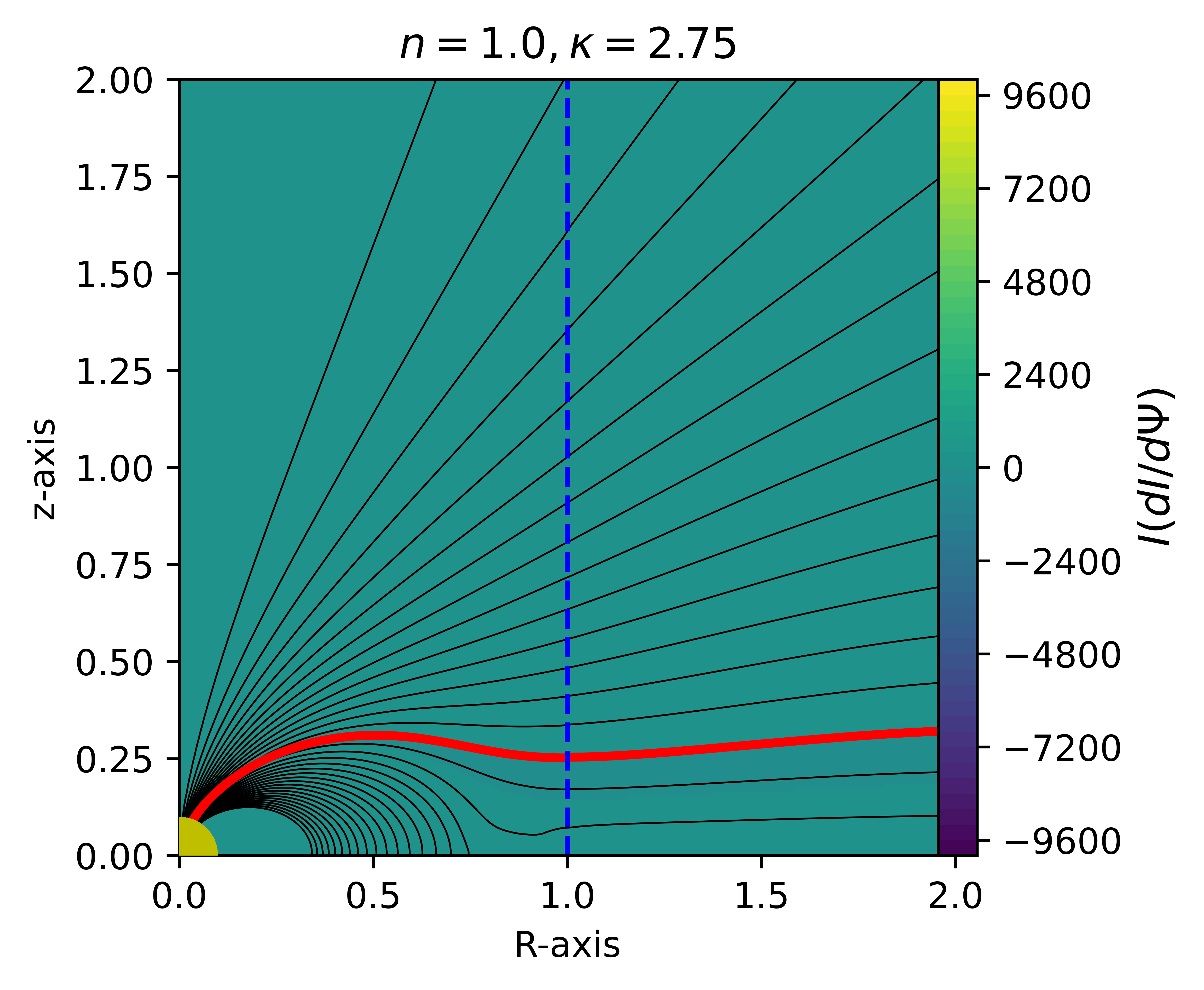}
    (b) \includegraphics[width=0.45\textwidth]{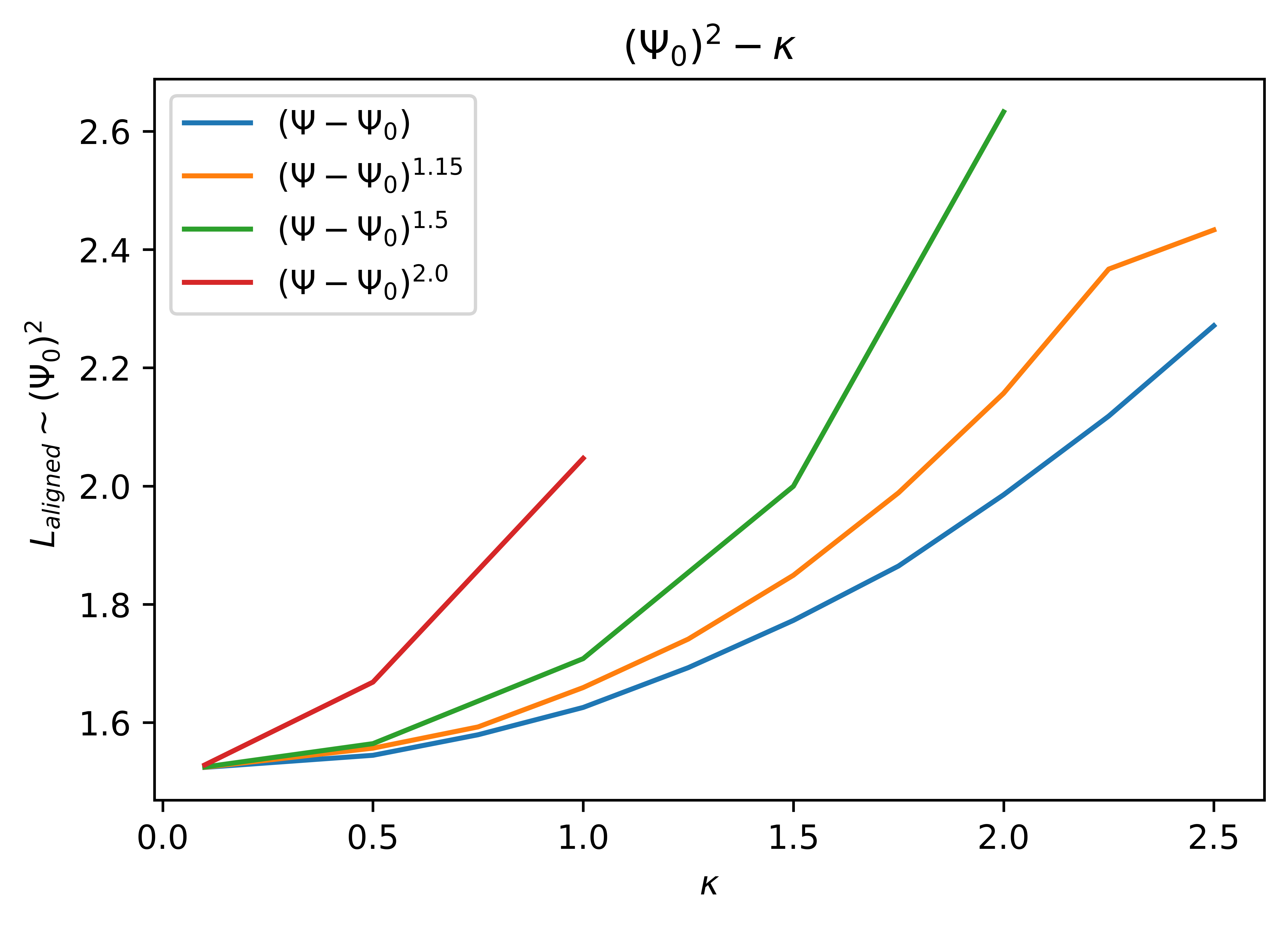}
    \caption{(a) Field inflation for a twisted magnetosphere with $\kappa=2.75$ and $n=1$, for a linear current function. The currents are shown in colour. The black lines are lines of constant flux,  the blue line is the light cylinder and the red line is the corresponds to the last open field line of the untwisted relativistic magnetosphere.    (b) The twist of the closed magnetic lines of the magnetosphere under the application of linear and non-linear current functions as a function of the numerical factor $\kappa$. As the current increases the twist increases and so do the open field lines. }.
    \label{fig:mesh1}
\end{figure*}
\unboldmath

Figure 1 (a) shows the magnetospheric structure in the presence of a poloidal current distribution in the closed magnetic lines with  $\kappa = 2.75\,, n = 1$. Note that the magnetic field line in red was the last closed field line of the untwisted solutions and it has become open and smoothly crosses the light cylinder. 
 We prescribe the twist of the closed field lines by the appropriate electric current function and we notice that by increasing the twist, a larger fraction of the magnetic field lines cross the light cylinder and become open, leading to large polar caps which may correspond to the hot spot during outburst phase. Our calculations reach a limiting twist about $\pi/2$ which is consistent with previous solutions (\cite{gourgouliatos2008fields} , \cite{Akgun:2017ggw}). If more twist is added, the field  becomes radial and prone to reconnection.
 Finally, we calculate the spin-down rate, using the relation $
  L \propto \Psi_0^2 $, 
where $L$ is the spin-down luminosity, and we find that it increases if more twist is added. It is possible that variations in the timing behavior of magnetars before and after outbursts can be associated to the magnetospheric twist. We conclude that by twisting the closed magnetic field lines the field becomes more radial, a larger fraction of magnetic flux crosses the light-cylinder and the spin-down rate becomes higher. Furthermore, we notice that the twist cannot be indefinitely increased, as a finite amount of shearing can lead to radial field lines. 

\bibliographystyle{123}

\bibliography{symposium_paper.bib}

\providecommand{\noopsort}[1]{}\providecommand{\singleletter}[1]{#1}%
\begin{thebibliography}{}

\bibitem[Akg\"un et~al., 2018]{Akgun:2017ggw}
Akg\"un, T., Cerd\'a-Dur\'an, P., Miralles, J.~A., and Pons, J.~A. (2018).
\newblock {The force-free twisted magnetosphere of a neutron star \textendash{}
  II. Degeneracies of the Grad\textendash{}Shafranov equation}.
\newblock {\em Mon. Not. Roy. Astron. Soc.}, 474(1):625--635.

\bibitem[Contopoulos et~al., 1999]{contopoulos1999axisymmetric}
Contopoulos, I., Kazanas, D., and Fendt, C. (1999).
\newblock The axisymmetric pulsar magnetosphere.
\newblock {\em The Astrophysical Journal}, 511(1):351.

\bibitem[{Goldreich} and {Julian}, 1969]{goldreich1969pulsar}
{Goldreich}, P. and {Julian}, W.~H. (1969).
\newblock {Pulsar Electrodynamics}.
\newblock {\em \apj}, 157:869.

\bibitem[Gourgouliatos and Lynden-Bell, 2008]{gourgouliatos2008fields}
Gourgouliatos, K. and Lynden-Bell, D. (2008).
\newblock Fields from a relativistic magnetic explosion.
\newblock {\em Monthly Notices of the Royal Astronomical Society},
  391(1):268--282.

\bibitem[Gourgouliatos and Lynden-Bell, 2019]{gourgouliatos2019coupled}
Gourgouliatos, K.~N. and Lynden-Bell, D. (2019).
\newblock Coupled axisymmetric pulsar magnetospheres.
\newblock {\em Monthly Notices of the Royal Astronomical Society},
  482(2):1942--1954.

\bibitem[{Lynden-Bell} and {Boily}, 1994]{1994MNRAS.267..146L}
{Lynden-Bell}, D. and {Boily}, C. (1994).
\newblock {Self-Similar Solutions up to Flashpoint in Highly Wound
  Magnetostatics}.
\newblock {\em \mnras}, 267:146.

\bibitem[{Parfrey} et~al., 2013]{2013ApJ...774...92P}
{Parfrey}, K., {Beloborodov}, A.~M., and {Hui}, L. (2013).
\newblock {Dynamics of Strongly Twisted Relativistic Magnetospheres}.
\newblock {\em \apj}, 774(2):92.

\bibitem[Tong, 2019]{10.1093/mnras/stz2438}
Tong, H. (2019).
\newblock {Large polar caps for twisted magnetosphere of magnetars}.
\newblock {\em Monthly Notices of the Royal Astronomical Society},
  489(3):3769--3777.

\bibitem[Vigan{\`o}, 2013]{vigano2013magnetic}
Vigan{\`o}, D. (2013).
\newblock Magnetic fields in neutron stars.
\newblock {\em arXiv preprint arXiv:1310.1243}.

\end{thebibliography}

\end{document}